\documentclass[twocolumn,showpacs,preprintnumbers,amsmath,amssymb,nofootinbib]{revtex4-1}

\usepackage{epsfig} 
\usepackage{graphicx}
\usepackage{dcolumn}
\usepackage{bm}
\usepackage{amsfonts}
\usepackage[usenames]{color}

\def\b{\begin{equation}}
\def\e{\end{equation}}
\def\be{\begin{eqnarray}}
\def\ee{\end{eqnarray}}

\newcommand{\insertplot}[5]{\begin{figure}
 \hfill\hbox to 0.05in{\vbox to #5in{\vfill
 \inputplot{#1}{#4}{#5}}\hfill}
 \hfill\vspace{-.1in}
 \caption{#2}\label{#3}
 \end{figure}}
 \newcommand{\inputplot}[3]{
 \special{ps: plotfile #1}
}
\newcounter{fig}

\begin{document}


\title{Kerr-Newman scalar clouds}

\author{Carolina L. Benone}
\email{lben.carol@gmail.com}
\affiliation{Faculdade de F\'{\i}sica, Universidade Federal do Par\'a, 66075-110, Bel\'em, Par\'a, Brazil}

\author{Lu\'{\i}s C. B. Crispino}
\email{crispino@ufpa.br}
\affiliation{Faculdade de F\'{\i}sica, Universidade Federal do Par\'a, 66075-110, Bel\'em, Par\'a, Brazil}

\author{Carlos Herdeiro}
\email{herdeiro@ua.pt}
\author{Eugen Radu}
\email{eugen.radu@ua.pt}
\affiliation{\vspace{2mm}Departamento de F\'\i sica da Universidade de Aveiro and I3N \\
Campus de Santiago, 3810-183 Aveiro, Portugal \vspace{1mm}}%

\date{\today}

\begin{abstract}
Massive complex scalar fields can form bound states around Kerr black holes. These bound states -- dubbed \textit{scalar clouds} -- are generically non-zero and finite on and outside the horizon; they decay exponentially at spatial infinity, have a real frequency and are specified by a set of integer ``quantum'' numbers $(n,l,m)$. For a specific set of these numbers, the clouds are only possible along a 1-dimensional subset of the 2-dimensional parameter space of Kerr black holes, called an \textit{existence line}. 
In this paper we make a thorough investigation of the scalar clouds due to neutral (charged) scalar fields around Kerr(-Newman) black holes. We present the location of the existence lines for a variety of quantum numbers, their spatial representation and compare analytic approximation formulas in the literature with our exact numerical results, exhibiting a sometimes remarkable agreement. 
\end{abstract}

\maketitle

\section{Introduction}
 
 The Schr\"odinger equation admits bound state solutions in a Coulomb potential. These are the atomic orbitals, familiar from elementary quantum mechanics. The corresponding scalar functions are finite everywhere and decay exponentially asymptotically. In the absence of spin, the orbitals can be labeled by three quantum numbers $(n,\ell,m)$, where $n$ counts the number of nodes of the radial function and $\ell,m$ are the standard spherical harmonic indices. The commonly used principal quantum number, which defines the orbital's energy, is $n+l+1$.
 
Changing this electromagnetic background to a black hole (BH) spacetime and the Schr\"odinger by the Klein-Gordon equation, one expects, at first sight, that no analogous scalar bound states should exist. Indeed, the causal structure of BH spacetimes demands that any classical field in the vicinity of the BH must be subjected to a purely ingoing boundary condition at the horizon. This seems to exclude equilibrium (i.e. stationary) configurations, and hence to rule out bound states around BHs. This expectation is actually confirmed for Schwarzschild BHs. Considering the massive Klein-Gordon equation $\Box \Psi=\mu^2\Psi$ in this background, in standard Schwarzschild coordinates, decomposing the solution into Fourier and harmonic modes, $\Psi\sim e^{-i\omega t} F(r)Y_{lm} (\theta,\phi) $, and requiring  these modes to have a radial exponential decay  as to describe bound states, $\lim_{r\rightarrow \infty} F(r)  \sim e^{-\sqrt{\mu^2-\omega^2}r}/r$,
one finds that the frequency $\omega$ must necessarily be complex. One finds, furthermore, that the imaginary part of the frequency, $\mathcal{I}(\omega)$, is always negative~\cite{Damour:1976kh,Zouros:1979iw,Detweiler:1980uk}; thus modes are decaying in time, and $\tau = 1/|\mathcal{I}(\omega)|$ measures the life-time of the decaying scalar configuration. The fact that $\mathcal{I}(\omega)\neq 0$  demonstrates the impossibility of an equilibrium between the scalar field and the BH, even if extremely long-lived configurations may exist~\cite{Barranco:2012qs}. This impossibility remains even if the backreaction of the minimally coupled scalar field is considered, i.e. at non-linear level within the Einstein-Klein-Gordon theory; this fact is confirmed by a number of no-(scalar)-hair theorems for spherically symmetric BHs~\cite{Bekenstein:1996pn,Pena:1997cy}.

 A remarkable change of affairs occurs when one considers Kerr BHs, for which the horizon is rotating with an angular velocity $\Omega_H$. Three qualitatively distinct types of massive scalar field modes which are asymptotically exponentially decaying can be found. For a mode with frequency $\omega$ and \textit{spheroidal} harmonic indices $l,m$ the imaginary part of the frequency is:
 
 
{\bf{(i)}} $\mathcal{I}(\omega)<0$, for $\mathcal{R}(\omega)>m\Omega_H$;  

{\bf{(ii)}} $\mathcal{I}(\omega)>0$, for $\mathcal{R}(\omega)<m\Omega_H$;  

{\bf{(iii)}} $\mathcal{I}(\omega)=0$, for $\omega=m\Omega_H$.  
 
 
 Regime ${\bf (i)}$ is the only one present for Schwarzschild BHs, as discussed above. Regime ${\bf (ii)}$ is called the \textit{superradiant regime}~\cite{Press:1972zz,Damour:1976kh,Zouros:1979iw,Detweiler:1980uk,Cardoso:2013krh,Shlapentokh-Rothman:2013ysa}; the corresponding scalar modes  can extract energy and angular momentum from the BH. It is made possible by the existence of an ergoregion. Regime ${\bf (iii)}$, i.e. when the scalar field frequency equals the \textit{critical frequency} $\omega_c\equiv m\Omega_H$, corresponds to bound states, analogous -- in terms of the scalar field distribution and labeling, but not in its probabilistic interpretation -- to the atomic orbitals. These are dubbed \textit{scalar clouds}~\cite{Hod:2012px,Hod:2013zza,Herdeiro:2014goa,Hod:2014baa}. 
 
Although scalar clouds have a phase-like time dependence, their energy-momentum tensor is time independent. As such, their spacetime backreaction is compatible with a stationary metric. Recently, the corresponding fully non-linear solutions of the coupled Einstein-Klein-Gordon system were found~\cite{Herdeiro:2014goa,Herdeiro:2014ima,Herdeiro:2014jaa}, corresponding to Kerr BHs with scalar hair.  These BHs can have quite distinct physical properties when compared to the canonical Kerr BHs and can lead to a different phenomenology, testable with future observations, such as gravitational wave observations~\cite{Hild:2011np,Okawa:2014nda,Degollado:2014vsa} and the Event Horizon Telescope~\cite{Loeb:2013lfa}. As such understanding their properties is timely. 

The goal of this paper is to perform a detailed study of neutral scalar clouds around Kerr BHs and charged scalar clouds around Kerr-Newman BHs, using a numerical approach, and comparing the results with some analytic formulas available in the literature. This study of scalar clouds is not only interesting from the viewpoint of BH theory, but it can also be seen as a step to understand the new type of hairy BHs we have mentioned.
 
 This paper is organized as follows. In Section \ref{seceq} we review the separation of variables procedure for solving the scalar wave equation in the Kerr-Newman background and the boundary conditions to be imposed in order to obtain bound state solutions. In Section \ref{seceq2} we will make a scan in the parameter space of Kerr and Kerr-Newman BHs to find the location of the existence lines of clouds for nodeless and nodeful clouds with different $(l,m)$ quantum numbers. Our results are obtained numerically, but we shall compare with  some analytic formulas existent in the literature obtained within some approximations. Then in Section \ref{sec3} we perform an analysis of the spatial distribution, both radial and angular, of a sample of clouds. We close with some final remarks in Section \ref{secfr}.

\section{Separation of variables: radial and angular equations}
\label{seceq}

We shall be considering a massive, charged scalar field minimally coupled to the geometry and to the electromagnetic potential of a rotating charged BH. The background spacetime is described by the Kerr-Newman line element in Boyer-Lindquist coordinates:
\be
{ds}^2&\ & = -\frac{\Delta}{\rho^2}(dt-a\sin^2{\theta}d\phi)^2 + \frac{\rho^2}{\Delta}dr^2\nonumber\\
 &+& \rho^2 d\theta^2 + \frac{\sin^2{\theta}}{\rho^2}[(r^2+a^2)d\phi-a dt]^2,
\ee
with
\b
\rho^2 \equiv  r^2 + a^2 \cos^2{\theta}, \hspace{0.5in} \Delta \equiv r^2 - 2Mr+a^2+Q^2,
\e
where $M$ and $Q$ are the ADM mass and charge of the BH, respectively, and the ADM angular momentum is given by $J=aM$. The background electromagnetic 4-potential is $A_\alpha=(-rQ/\rho,0,0,aQr \sin^2{\theta}/\rho)$.

The Klein-Gordon equation for a massive charged particle is given by
\b
(\nabla^\alpha-iqA^\alpha)(\nabla_\alpha-iqA_\alpha) \Psi - \mu^2 \Psi = 0,
\e 
where $\mu$ is the mass of the scalar field and $q$ is its charge. In order to solve this equation we decompose the scalar field as $\Psi=\sum_{l,m}\Psi_{lm}$ and separate variables as $\Psi_{lm} =R_{lm}(r)S_{lm}(\theta)e^{im\phi}e^{-i\omega t}$~\cite{Brill:1972xj}, where $S_{lm}(\theta)$ are the spheroidal harmonics which obey

\begin{widetext}
\b
\frac{1}{\sin{\theta}}\frac{d}{d\theta}\left(\sin{\theta}\frac{d S_{lm}}{d \theta}\right)+\left(K_{lm}+a^2(\mu^2 - \omega^2) -
a^2(\mu^2 - \omega^2)\cos{\theta}-\frac{m^2}{\sin^2{\theta}}\right)S_{lm} =0 \ .
\label{esh}
\e
$K_{lm}$ are separation constants. The radial functions $R_{lm}(r)$ then obey the radial equation
\b
\Delta \frac{d}{d r}\left(\Delta \frac{d R_{lm}}{d r}\right) + \left[H^2 + (2ma\omega - K_{lm} - \mu^2 (r^2+a^2))\Delta \right]R_{lm}=0,
\label{eqr}
\e
\end{widetext}
where $H\equiv (r^2+a^2)\omega-am-qQ r$. We can rewrite Eq. (\ref{eqr}) using the tortoise coordinates, defined by
\b
\frac{dr_*}{dr} \equiv  \frac{r^2+a^2}{\Delta} \ ,
\label{tor}
\e
and obtain a new radial equation without the first derivative term,
\begin{widetext}
\b
\frac{d^2 U_{lm}}{dr_*^2} + \left\{\frac{[H^2+ (2ma\omega-\mu^2 (r^2+a^2) - K_{lm})\Delta]}{(r^2+a^2)^2}-\frac{\Delta(\Delta+2r(r-M))}{(r^2+a^2)^3}+ \frac{3r^2\Delta^2}{(r^2+a^2)^4}\right\}U_{lm}=0 \ ,
\label{ert}
\e
\end{widetext}
in terms of the new dependent functions $U_{lm}$ defined as 
\b
U_{lm}\equiv R_{lm}\sqrt{r^2+a^2} \ . 
\e

Next we must impose boundary conditions. Any state in a BH background should have a purely ingoing boundary condition at the horizon; moreover the bound states we are looking for should have an asymptotically exponentially decaying behavior.  Analyzing the radial equation (\ref{eqr}) we find asymptotic solutions compatible with these requirements
\b
R_{lm}(r) \approx 
\left\{ 
\begin{array}{ll}
e^{-i(\omega-\omega_c) r_*}, \quad &\mbox{for $r\rightarrow r_+$},\\
 \frac{e^{-\sqrt{\mu^2-\omega^2} r}}{r}, \quad &\mbox{for $r\rightarrow \infty$},
\end{array}
\right.
\label{sol}
\e
where we defined the critical frequency $\omega_c$, given by
\b
\omega_c \equiv m \Omega_H + q\Phi_H = \frac{ma}{r_+^2+a^2}+ \frac{qQr_+}{r_+^2+a^2} \ .
\label{frequency_horizon}
\e
Here, $r_+$ is the radial coordinate of the outer (event) horizon,  $\Omega_H$ is the horizon angular velocity, as mentioned in the  Introduction,  and $\Phi_H$ is the horizon electric potential. Later we shall also use $r_-$, the radial coordinate of the inner (Cauchy) horizon. Observe that in the presence of both background and scalar field electric charge, the critical frequency gets a new contribution, as compared to that discussed in the Introduction. Indeed there is a purely charged superradiance that, under appropriate boundary conditions, can also lead to instabilities \cite{Herdeiro:2013pia,Hod:2013fvl,Degollado:2013bha}.

\section{Scanning the BH parameter space for clouds}
\label{seceq2}
In order to study bound states, in the following, we shall focus on the case for which the field's frequency equals the critical one 
\b
\omega=\omega_c \ .
\label{cond}
\e
 This choice allows the existence of stationary scalar  configurations around Kerr-Newman BHs, but only for specific values of the background parameters; mathematically, one may regard \eqref{ert} as a non-standard eigenvalue problem. In our approach, these specific values will be found numerically. Our strategy can be summarized
  as follows. 
  The radial equation (\ref{eqr}) is solved for given
cloud quantum numbers $(n,l,m)$ and charge $q$. 
The field mass $\mu$ is taken as a normalization scale and all quantities will be referred with respect to it. 
Moreover, we fix the BH background parameters $r_+$ and $Q$. 
 In this procedure, we consider the following expansion for the coupling constant $K_{lm}$
\b
K_{lm} + a^2(\mu^2-\omega^2) = l(l+1)+\sum_{k=1}^\infty c_k a^{2k}(\mu^2-\omega^2)^{k},
\e
where the coefficients $c_k$ may be found in Ref. \cite{abramowitz+stegun}.

As $r\to r_+$, the radial function (with the critical frequency (\ref{cond}))
admits a power series expansion,  
\begin{eqnarray}
\label{near-horizon}
R_{lm}=R_0\bigg(1+\sum_{k\geq 1}R_k(r-r_+)^k\bigg),
\end{eqnarray}
 with $R_0$ an arbitrary nonzero constant 
 (since we consider only the linear Klein-Gordon equation).
 The coefficients $R_k$ are found by replacing Eq.~(\ref{near-horizon}) into the Eq.~(\ref{eqr}),
 and solving it order by order in terms of $(r-r_+)$.
 In our numerics, we have considered only the 
 $k=1,2$ terms in (\ref{near-horizon})
 and took, without loss of generality, $R_0=1$. 
The $R_k$ exhibit a non-elucidating dependence 
on the background parameters $(r_+,a,Q)$, on $q$ 
and on the quantum numbers $(l,m)$; thus we shall not exhibit them here.

 Then, starting with the near horizon expansion (\ref{near-horizon}),
  we search for values of $a$ for which the radial function $R_{lm}$ 
 goes to zero (exponentially) at infinity, as given by the second relation
 in Eq.~(\ref{sol}).
The numerical integration of Eq.~(\ref{eqr}) 
results in a one-parameter shooting problem,
which was solved
using both a standard  \textsc{fortran} solver,
as well as a \textsc{mathematica} routine, with  agreement to high accuracy.
We have found that, for given input parameters  $(r_+,Q,q;l,m)$,
solutions with the right asymptotics
exist for a discrete set of $a$,
which can be labeled by the number $n$
of nodes of the radial function $R_{lm}$. 
In this way we determine the existence lines of the clouds with a given set of quantum numbers, in the parameter space of Kerr-Newman BHs.

We shall now present the results obtained numerically for the clouds. Firstly, we consider the case of a massive scalar field in the Kerr spacetime ($Q=0$ and $q=0$); then we discuss the case of a massive charged scalar field in the Kerr-Newman spacetime. We always assume the cosmic censorship hypothesis, so that the singularities are hidden by event horizons; in other words, we never consider over-extreme backgrounds.

\subsection{Kerr}
For Kerr BHs, the scalar field critical frequency ($\omega_c$) and the background horizon angular velocity are related by $\Omega_H=\omega_c/m$. The existence lines for nodeless clouds and $l=m=1,2,3,4,10$ were first exhibited in~\cite{Herdeiro:2014goa}, using a BH mass $M$ vs. horizon angular velocity $\Omega_H$ diagram. This particular type of diagram parameterizes the 2-dimensional parameter space of Kerr BHs in a way appropriate for this problem, due to the relation between the scalar field frequency and the horizon angular velocity. As such, we shall herein represent existence lines using the same type of diagram. 

\begin{figure}[h!]
\centering
\includegraphics[height=2.5in]{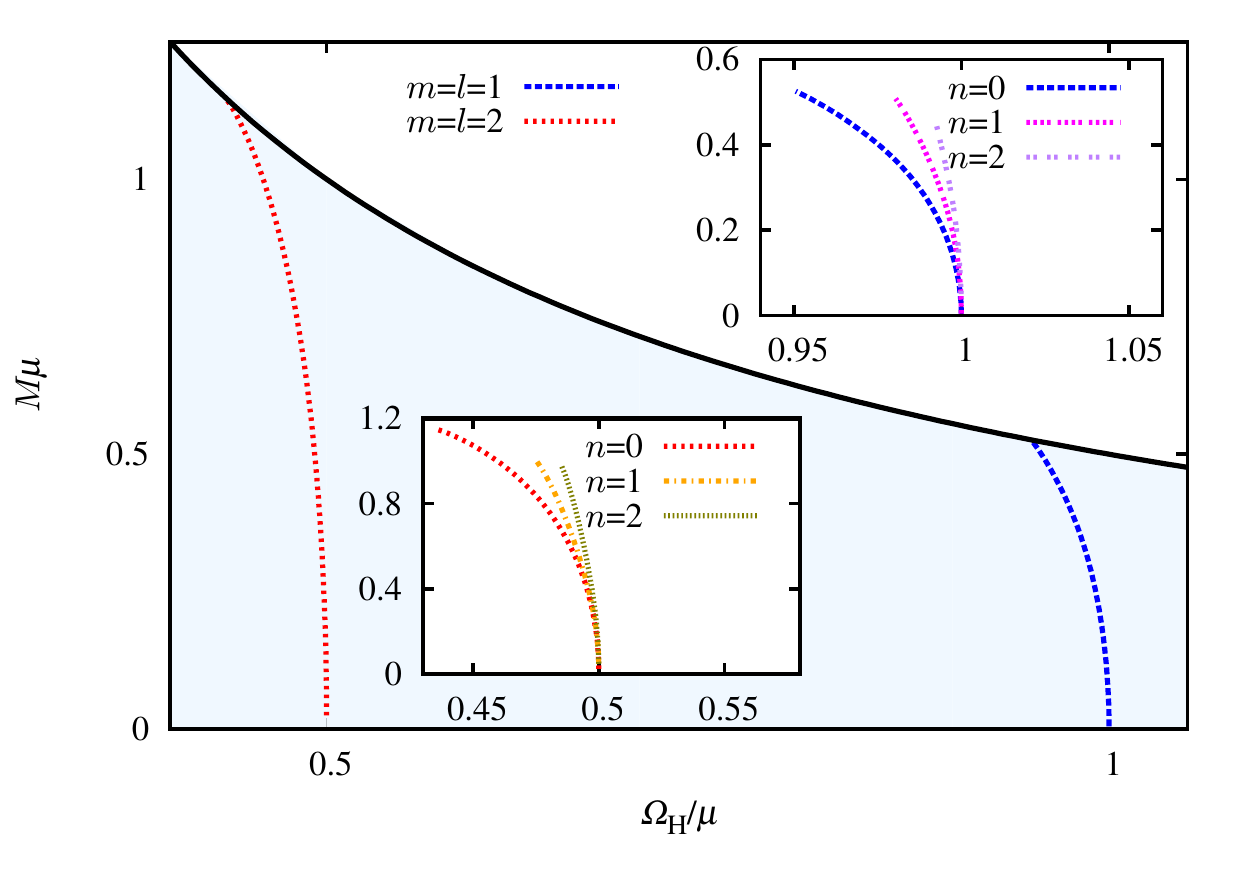}
\caption{Existence lines for scalar clouds with various quantum numbers in the mass vs. horizon angular velocity parameter space of Kerr BHs. The black solid curve represents the extreme case, $a=M$ and Kerr solutions exist below this line. The blue dashed and red dotted lines represent the nodeless solution for $m=l=1$ and $m=l=2$, respectively. The insets compare the nodeless solutions ($n=0$) with the solutions with $n=1,2$.} 
\label{fel}
\end{figure}

In Fig. \ref{fel} we plot the existence lines for the clouds with different node numbers, $n=0,1,2$, and angular momentum harmonic indices, $l=m=1,2$. As mentioned before, the lines with $n=0$ and $l=m=1,2$ have already been presented in Ref.~\cite{Herdeiro:2014goa}. The main trend concerning these lines is that as $l=m$ is increased the lines move towards smaller $\Omega_H$. The main new feature presented here is that  the solutions with nodes move towards larger values of $\Omega_H$ as compared to nodeless solutions with the same $l,m$, converging to the latter when $M\rightarrow 0$. In Section~\ref{secfr} we shall provide an intuitive interpretation for the behaviour of these and the following existence lines.

\begin{figure}[h!]
\centering
\includegraphics[height=2.5in]{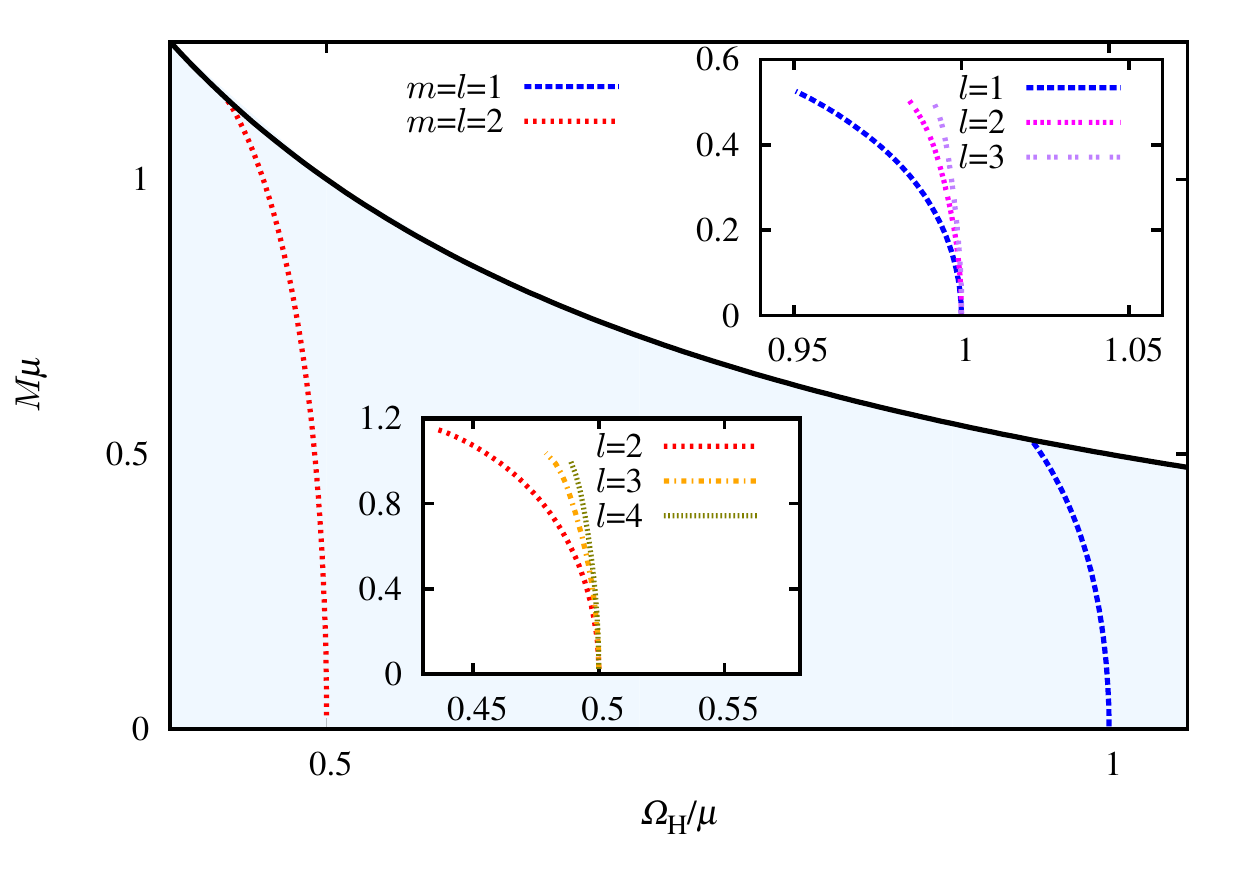}
\caption{Analogous plot to Fig.~\ref{fel}, but now the insets compare the solutions for $m=l$ with the solutions with $m < l$, all with $n=0$.} 
\label{felm}
\end{figure}

It was briefly mentioned in Ref.~\cite{Herdeiro:2014goa} that the existence lines for nodeless clouds with a given value of $m$ and $l>m$ always move towards larger $\Omega_H$ values than the corresponding ones with $m=l$. This is illustrated in Fig. \ref{felm}. The trend we have seen in Fig. \ref{fel} for the existence lines with nodes establishes a similar pattern when $l,m$ are fixed and we increase $n$. As such, fixing $m$, the existence line for any $n,l$ that stands on lowest values of $\Omega_H$ is the $n=0$, $l=m$ line. We recall that the region to the right of a given existence line -- in this type of diagram -- consists of background spacetimes that are superradiantly unstable against that particular mode. Consequently, the $m=l$, $n=0$ existence line defines  the boundary of the region between stable and unstable Kerr solutions for a given $m$ mode.

Although there is no general analytic formula for the clouds' existence lines, some limiting cases have been considered in the literature which led to analytic formulas valid within some approximation. Here we shall discuss two such limits, one that applies to fast rotating BHs and another that applies to slowly rotating BHs. 

In Ref.~\cite{Hod:2012px} Hod first discussed the clouds for the extremal Kerr BH and in Ref.~\cite{Hod:2013zza} extended his results for near-extremal BHs, obtaining an analytic approximation given by (cf. Eq. (32) in Ref.~\cite{Hod:2013zza}) 

\begin{figure}[h!]
\centering
\includegraphics[width=\columnwidth]{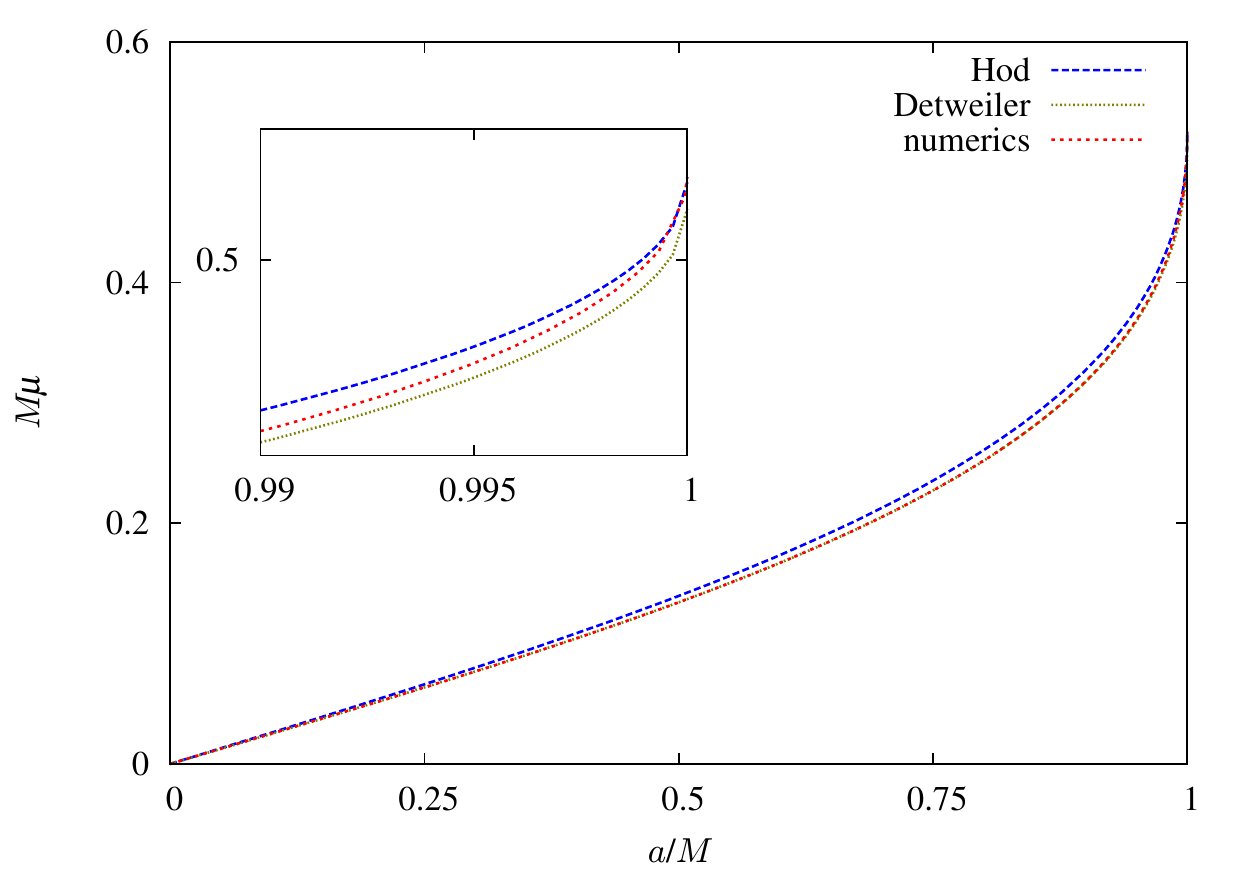}
\caption{Comparison between our numerical solution for the clouds with $n=0$, $m,l=1$ and the analytical results by Hod~\cite{Hod:2013zza}, cf. Eq.~\eqref{hap}, and Detweiler~\cite{Detweiler:1980uk}, cf. Eq.~\eqref{dap}, in a mass vs. angular momentum parameter for Kerr BHs.}
\label{nhd}
\end{figure}

\b
\mu = m \Omega_H[1+2\bar{\epsilon}^2] \,,
\label{hap}
\e
where
\b
\bar{\epsilon} = \frac{m}{2(d+1+2n)} - \frac{m^3}{4d(d+1+2n)^2}\left(\frac{r_+-r_-}{r_+}\right)
\e
and $d = \sqrt{(2l+1)^2-4m^2}$. From these equations we can obtain the existence line for a given cloud with quantum numbers $(n,l,m)$ and for \textit{rapidly} rotating BHs, i.e. near extremality. In Fig.~ \ref{nhd}, however, we plot such line for $n=0$, $l=m=1$ in an $M$ vs. $a/M$ diagram for Kerr, but \textit{extrapolating} the formula for all values of $a/M$.

\begin{figure}[h!]
\centering
\includegraphics[width=\columnwidth]{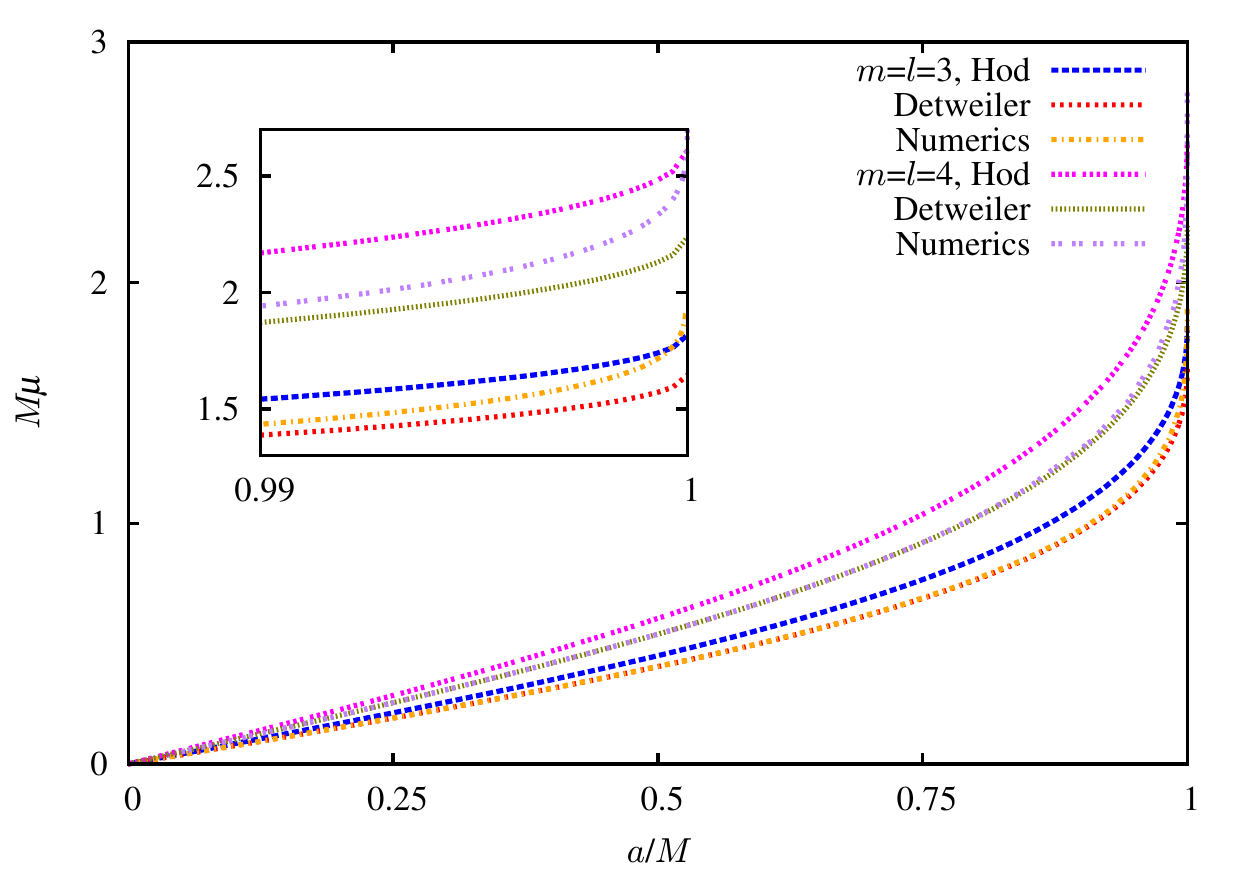}
\caption{Analogous comparison to that in Fig. \ref{nhd} but now for the clouds with $n=0$, $m=l=3,4$. The agreement between the analytic and numerical approximations seems to become slightly worse, outside their regime of validity, when the quantum numbers $m=l$ increase.}
\label{nhdm}
\end{figure}

Another analytic formula can be obtained from the classic work of Detweiler~\cite{Detweiler:1980uk}, who studied superradiance for small values of the mass coupling, $M\mu \ll 1 $. His results can be specialized to the critical frequency $\omega_c=m\Omega_H$. Since $\omega<\mu$ for bound states, we obtain that Detweiler's results apply to \textit{slowly} rotating BHs, i.e. $\Omega_H M\ll 1$; then, an analytical formula can be obtained (solving Eq. (26) of Ref.~\citep{Detweiler:1980uk} for $\mu$), namely:
\b
\mu=\frac{1}{\sqrt{2}}\left[\frac{p^2}{M^2}-\frac{\sqrt{p^2 (p^2-4 M^2 m^2 \Omega^2)}}{M^2}\right]^{1/2},
\label{dap}
\e
where $p=l+n+1$. In Fig. \ref{nhd} we plot the corresponding existence line for $l=m=1$ and $n=0$, \textit{extrapolating} to all values of $a$. In this figure, besides the approximations derived from Eq.~\eqref{hap} and Eq.~\eqref{dap} we plot our numerical results, which are valid (and accurate) for all values of $a/M$. Somewhat unexpectedly, the analytic approximations are still accurate, well outside their regime of validity. Thus, even though the solution of Detweiler is supposed to be valid only in the slowly rotating case, we see that the numerical and the Detweiler curves overlap in almost all the range for $a/M$. In the inset it is possible to see that the numerical result tends to the Hod curve close to extremality, as expected. In Fig. \ref{nhdm} a similar comparison is made for $n=0$ and $m=l=3,4$.

\subsection{Kerr-Newman}
\begin{figure}[h!]
\centering
\includegraphics[height=2.5in]{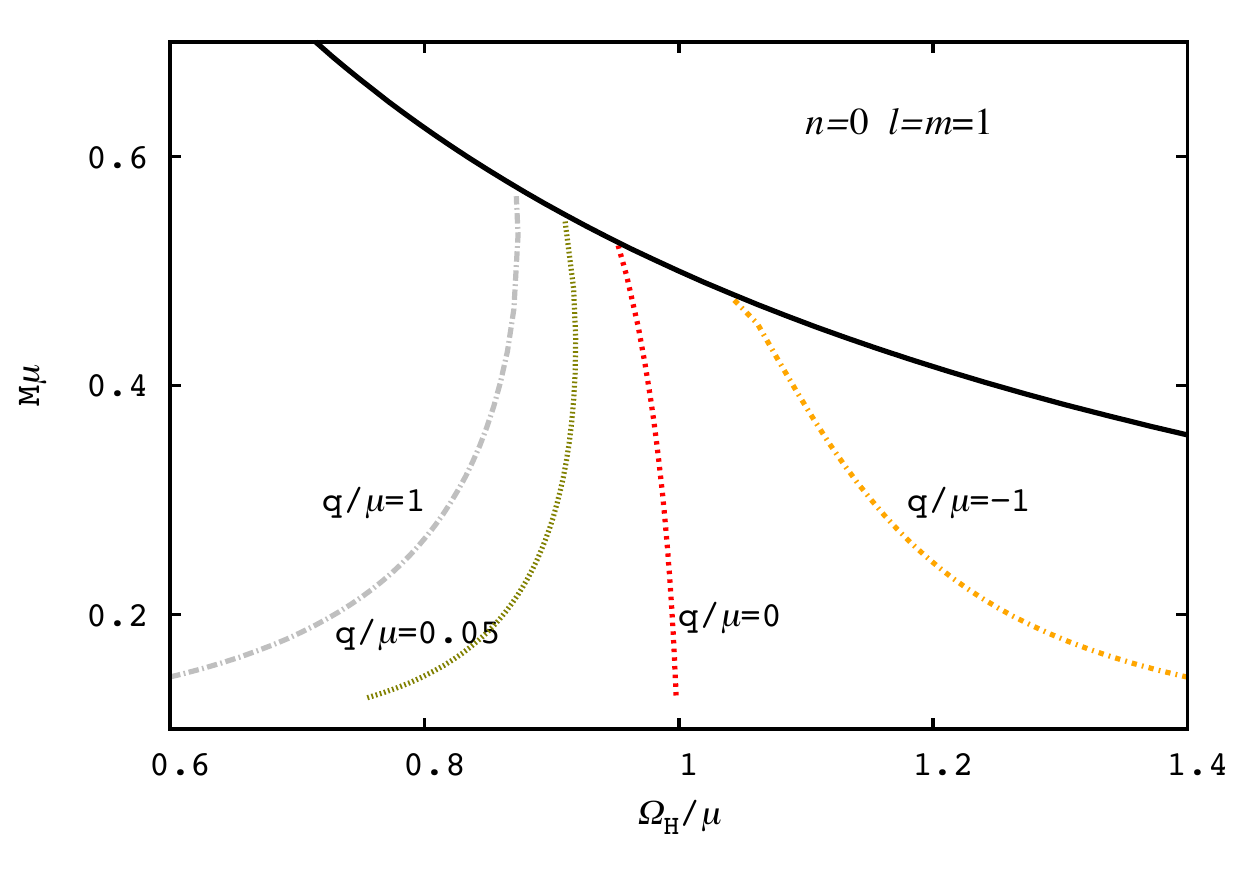}
\caption{Existence lines for charged scalar bound states in the Kerr-Newman background, for $n=0$ and $l=m=1$, for different values of the field charge and fixed background charge $\mu Q=0.1$.} 
\label{fkn1}
\end{figure}

In the Kerr-Newman case both the background and the test field have one more parameter. So to exhibit the existence lines in a useful way, one must fix some quantities. In Fig. \ref{fkn1} we fix the background charge $\mu Q=0.1$ and draw the existence lines for the cloud with $l=m=1$ and $n=0$ for various values of the field charge $q$. The overall trend is that clouds with the same (opposite) charge as the background occur for smaller (larger) angular velocities. This is an intuitive behavior. For instance, same charge implies Coulomb repulsion and hence require a smaller angular velocity from the background to maintain the equilibrium.

\begin{figure}[h!]
\centering
\includegraphics[height=2.5in]{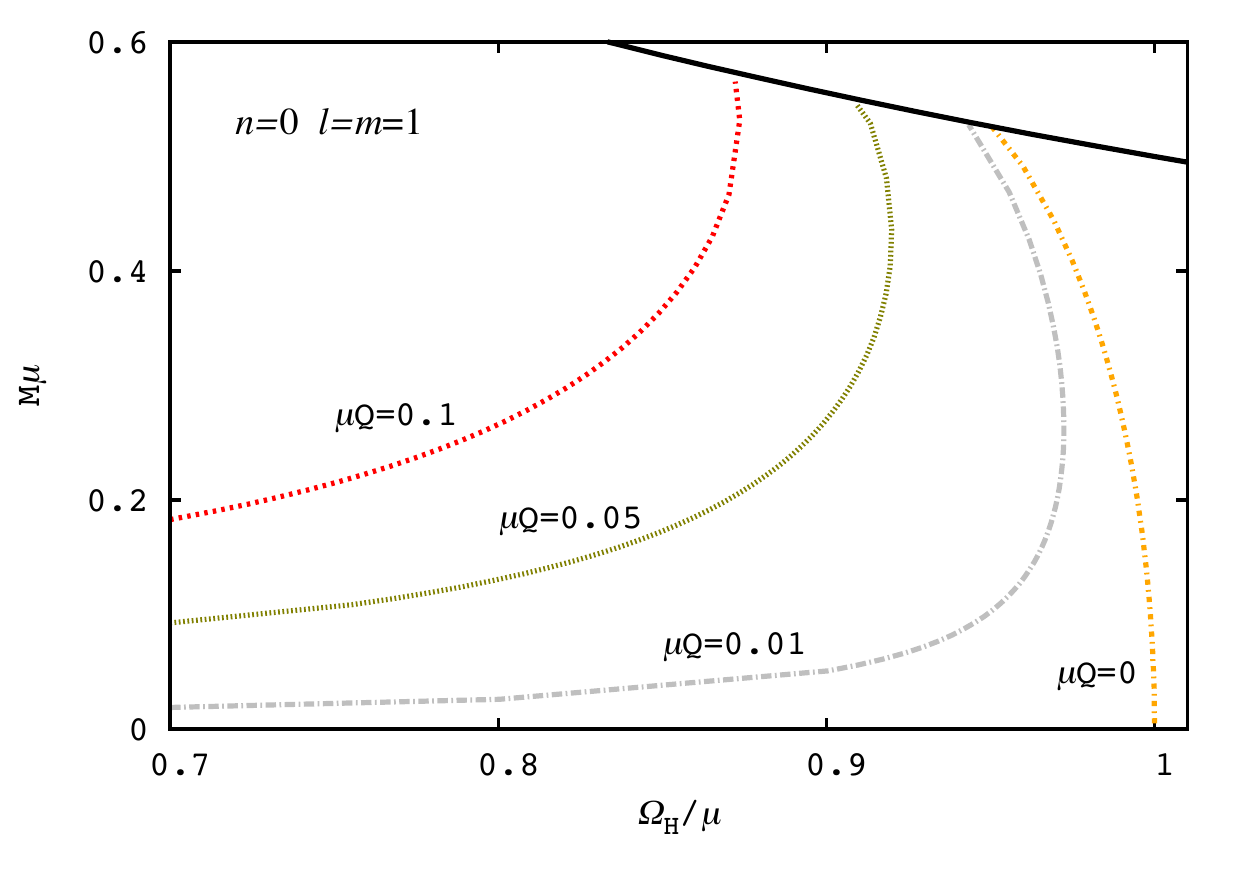}
\caption{Existence lines for charged scalar bound states in the Kerr-Newman background, for $n=0$ and $l=m=1$, for different values of the background charge and fixed field charge $q/\mu=1$.} 
\label{fkn2}
\end{figure}

\begin{figure}[h!]
\centering
\includegraphics[height=2.5in]{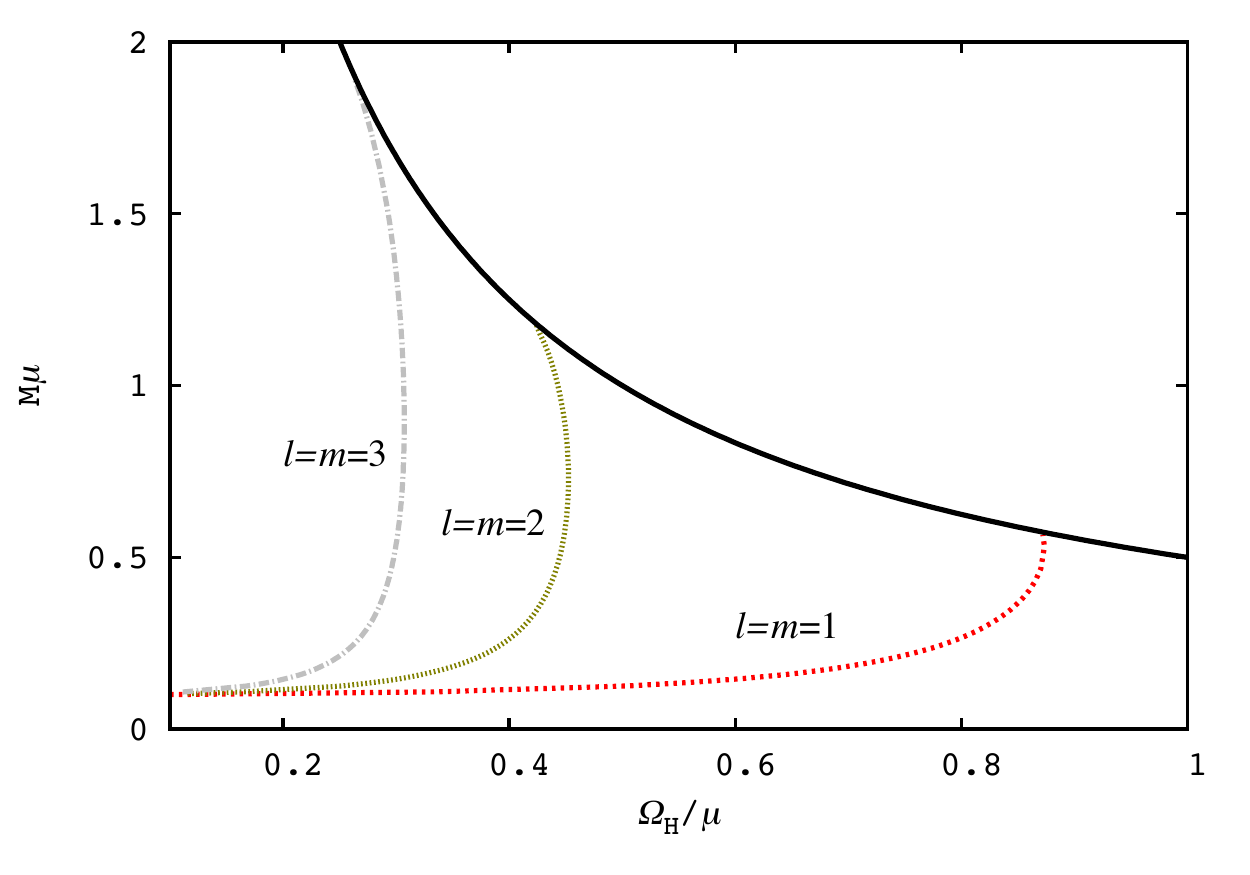}
\caption{Existence lines for charged scalar bound states in the Kerr-Newman background, for $\mu Q=0.1$, $q/\mu=1$, $n=0$ and $m=l=1,2$ and $3$.} 
\label{fkn3}
\end{figure}

Another distinct feature of the charged existence lines, already seen in Fig.  \ref{fkn1} but more clearly exhibited in Figs.  \ref{fkn2} and  \ref{fkn3}  is that the existence lines do not reach $M=0$, since the inclusion of background charge implies a minimum value for the background mass, i.e. $|Q|<M$. Moreover,  Fig.  \ref{fkn2} confirms the trend that increasing the Coulomb repulsion between the field and the background implies that for the same background mass the clouds exist for lower background angular velocity.

Finally, fixing both the background and field charge, the variation of the existence lines when the field's angular momentum quantum numbers are increased is qualitatively similar to that seen in the Kerr case, namely the lines move towards lower angular velocities, as it can be seen in Fig.~\ref{fkn3}. This may be interpreted as a trade off between the angular momentum of the background and that of the cloud, as to maintain equilibrium. Another interesting feature in Fig.~\ref{fkn3} is that, as the minimum background mass is approached, the angular velocity of the background tends to zero. Observe that the minimum background mass is precisely equal to the charge $\mu Q=\mu M=0.1$ and that the field mass and charge are also the same $q/\mu=1$. Thus, the limiting equilibrium configuration is a \textit{marginal (charged) cloud} of the type discussed in Refs.~\cite{Degollado:2013eqa,Sampaio:2014swa}.

\begin{figure}[t!]
\centering
\includegraphics[height=2.5in]{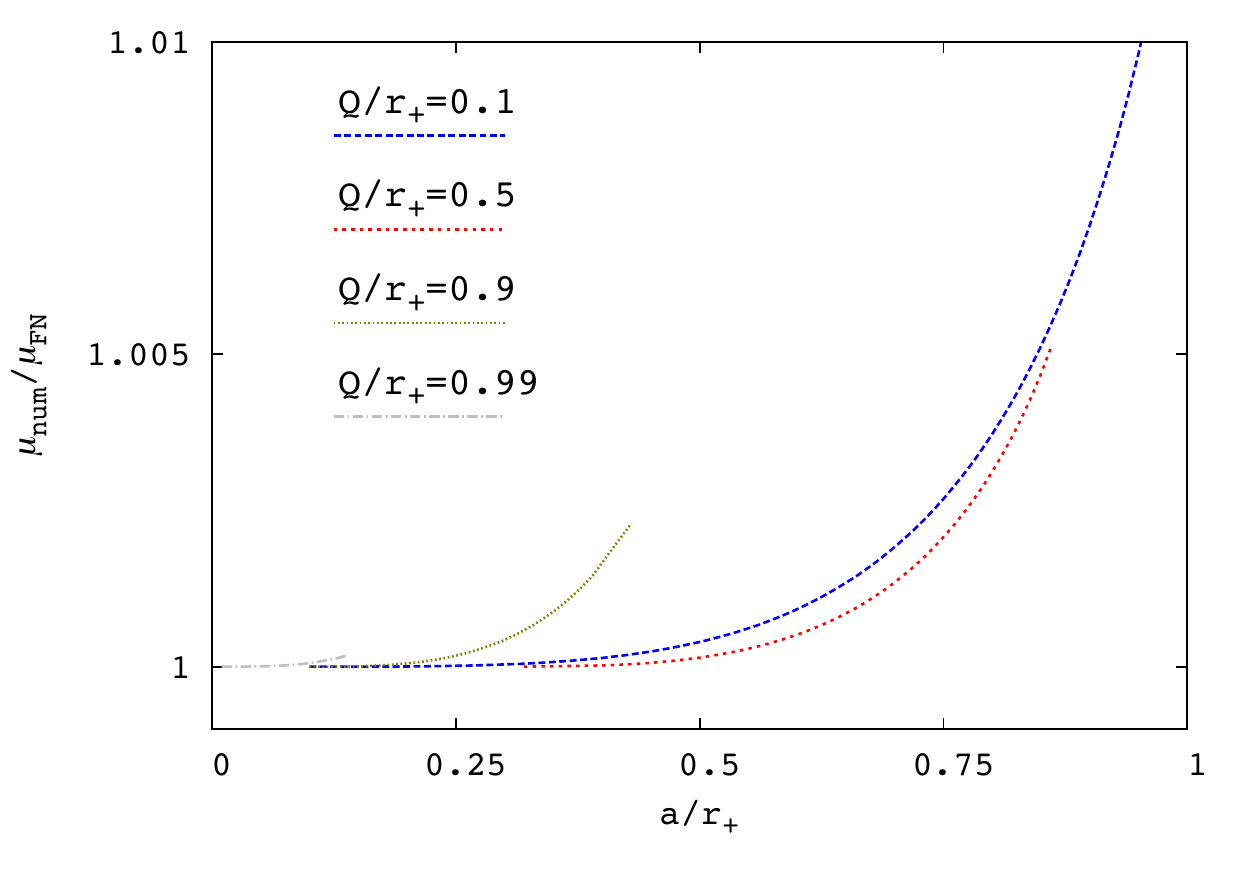}
\caption{Comparison between our numerical solutions and the analytical formula by Furuhashi and Nambu~\cite{Furuhashi:2004jk}, cf. Eq.~\eqref{fap}, for clouds with $n=0$, $m=l=1$ and $q r_+=0.1$. The numerical and analytical formulas coincide with very good accuracy.} 
\label{fkn}
\end{figure}

As for the Kerr case, we can compare our numerical results for the Kerr-Newman case with an analytic formula. The latter was obtained from the results of Furuhashi and Nambu~\cite{Furuhashi:2004jk}. First we note that these authors have shown that in order to have bound states we must have
\b
M\mu \gtrsim Qq \ .
\label{muq}
\e
Then they obtained an expression for the real part of the frequency for $M\mu \ll 1$ and $Qq\ll 1$ (cf. Eq. (26) of Ref.~\cite{Furuhashi:2004jk}). From  that expression, we find the analytic formula
\b
\mu=\mathcal{R}\left[\frac{2 qQ}{3 M}+\frac{(1-i\sqrt{3})(6p^2 +q^2Q^2)M^2}{2^{2/3}3M^2 A}+\frac{(1+i\sqrt{3} A)}{2^{1/3}6M^2 }\right],
\label{fap}
\e
where
\be
A &=&\{-36 p^2 M^3 q Q+2 M^3 q^3 Q^3+54 g^2 M^4 \omega_c\nonumber\\
&+&[4 (-6 p^2 M^2-M^2 q^2 Q^2)^3\nonumber\\
&+&4M^6(-18 p^2 q Q+q^3 Q^3+27 p^2 M \omega_c)^2]^{1/2}\}^{1/3}.
\ee
In Fig. \ref{fkn} we compare this analytic formula with our numerical results and, again, conclude that the analytic formula works remarkably well. Observe that in the Kerr-Newman case the existence lines cannot extend in the whole range of $a$, since they are constrained by the conditions (\ref{muq}) and $a^2+Q^2<M^2$.

\section{Cloud tomography}
\label{sec3}
We will now consider the spatial distribution of clouds. All considerations in the following will be made using the standard Boyer-Lindquist coordinates for the Kerr-Newman spacetime. Since the angular and radial dependence separate for each cloud with fixed $(n,l,m)$ it suffices to consider these dependences separately to obtain the full spatial picture.

\subsection{Angular functions}
\label{seceq21}

The angular dependence is given by spheroidal harmonics. These harmonics depend on the background angular momentum parameter $a$; $S_{lm}e^{im\phi}$ reduce to the standard spherical harmonics $Y_{lm}$ when $a=0$, up to a $(l,m)$-dependent normalization factor. Since one is typically less familiar with these spheroidal harmonics (than with the spherical harmonics) we will illustrate their angular distribution.

In Fig.  \ref{slm3} we give a 3-dimensional plot of some spheroidal harmonics for $\mu r_+=0.5$ 
As we increase the value of $l=m$, $S_{lm}$ becomes more flattened, as for spherical harmonics. 
For the cases plotted, the difference between spherical and spheroidal is essentially only an overall scale factor, i.e. the angular distribution is very similar.  Obviously the angular dependence is independent of considering the Kerr or the Kerr-Newman background.

\subsection{Radial functions}
\label{seceq22}
The radial dependence of the clouds is quite simple and is illustrated in Figs. \ref{3dergo} (for Kerr) and \ref{rkn} (for Kerr-Newman), for some values of the rotation parameter, for two values of $l=m$ and for three different numbers of nodes $n$. The scalar field is always finite on and outside the horizon. On the horizon it needs not  be zero. For instance, for $l=m=1$ clouds it is non-zero on the horizon, cf. Figs. \ref{3dergo} and \ref{rkn}. Then, the radial function will have $n$ nodes and decrease exponentially asymptotically. 

\begin{widetext}

\begin{figure}[h!]
\centering
\includegraphics[height=1.8in]{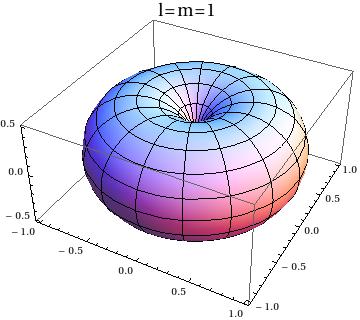}
\includegraphics[height=1.8in]{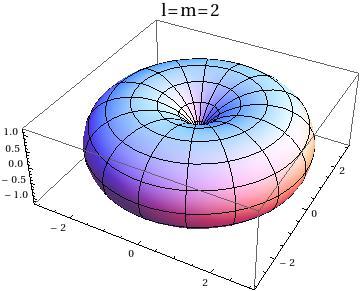}
\includegraphics[height=2in]{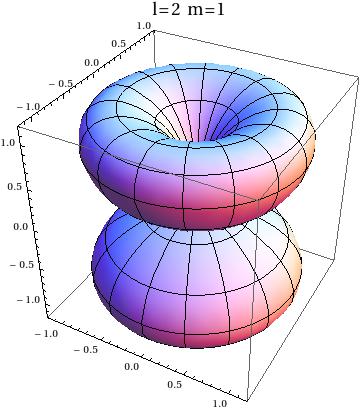}
\caption{3-dimensional plots of the spheroidal harmonics $|S_{lm}(\theta)|$. All three panels were obtained for $\mu r_+=0.5$. These correspond, from left to right to $ \mu a=0.399$, $\mu a=0.133$ and $\mu a=0.430$.}
\label{slm3}
\end{figure}

\begin{figure}[h!]
\centering
\includegraphics[height=2.2in]{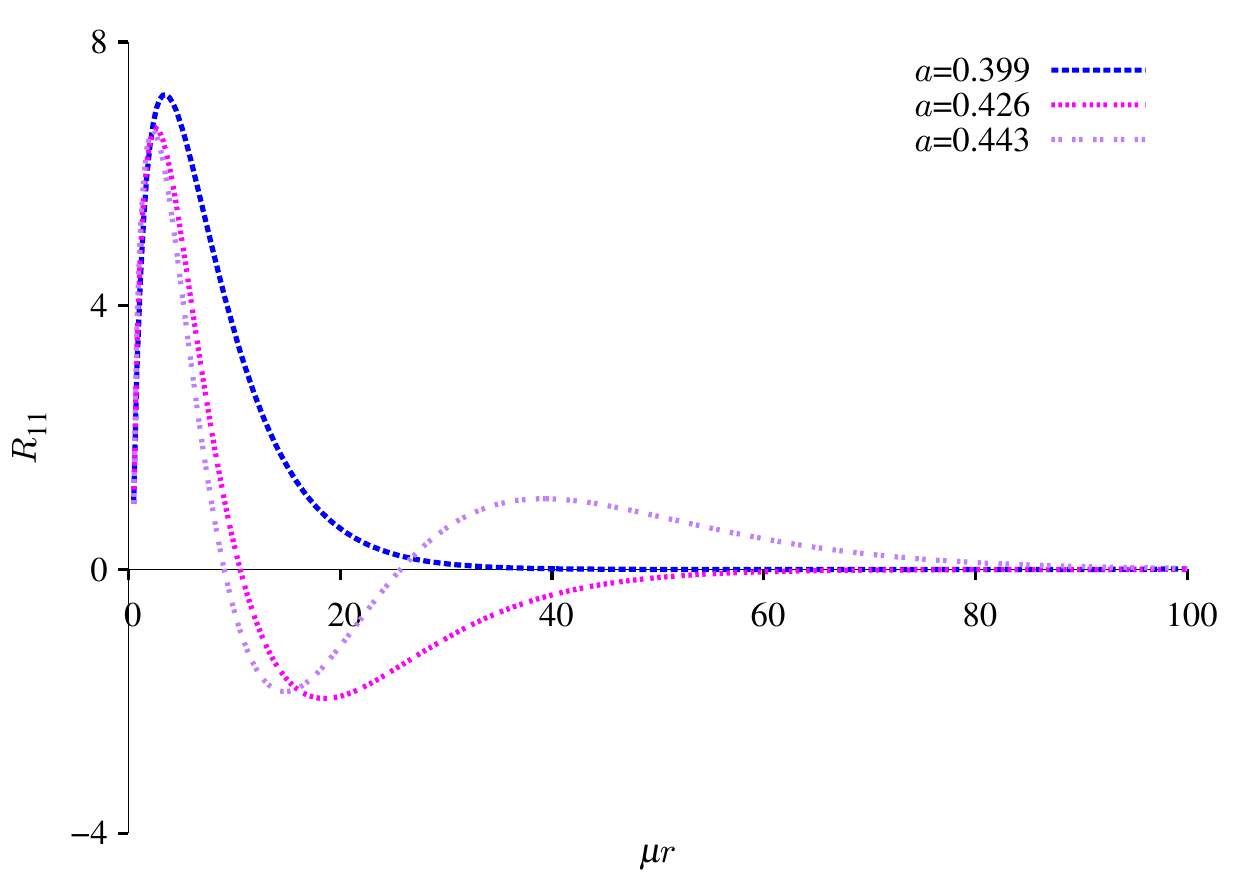} \ \ \ 
\includegraphics[height=2.2in]{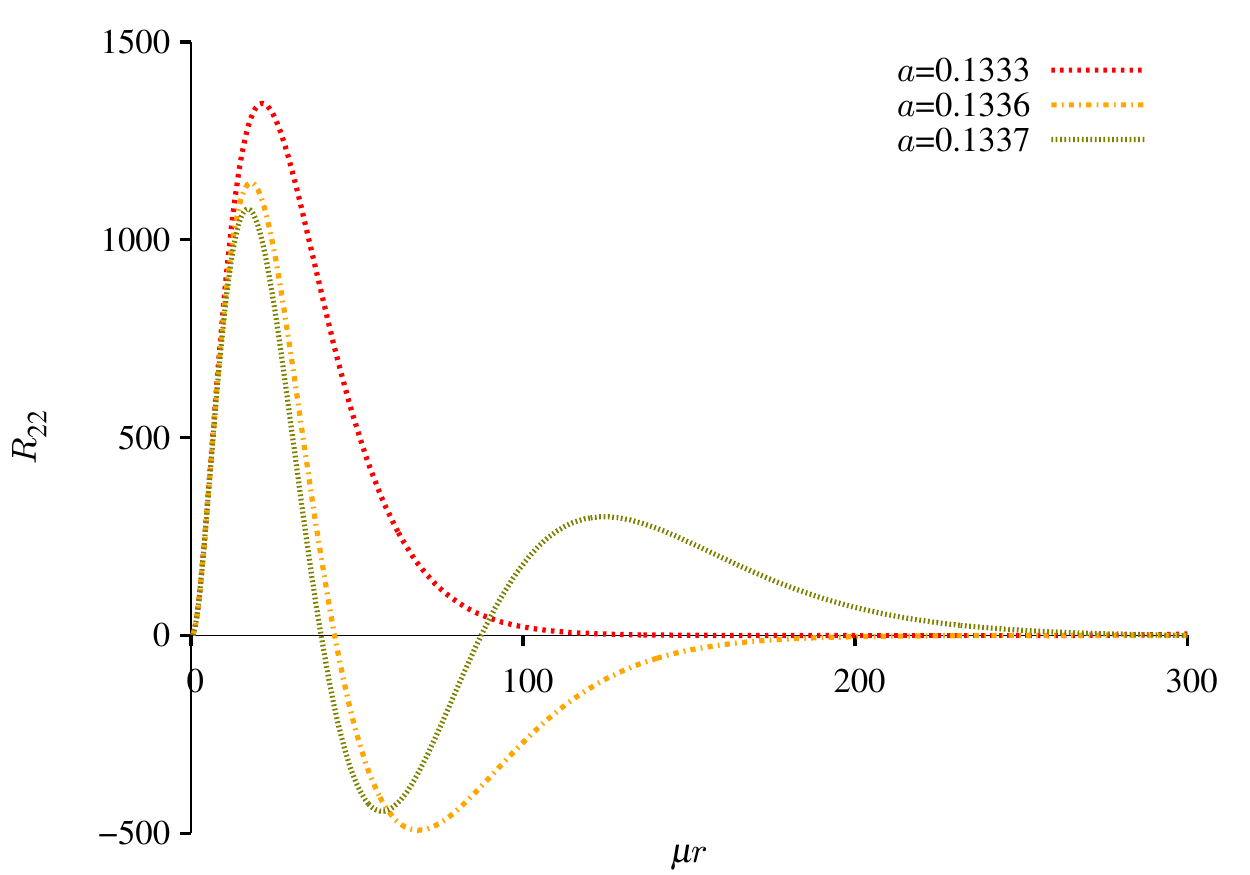}
\caption{Radial solutions $R_{11}$  (left), $R_{22}$ (right) for clouds with $n=0,1,2$ in the Kerr background with $r_+=0.5$. The corresponding values of $a$ are given in the figure key.} 
\label{3dergo}
\end{figure}
\begin{figure}[h!]
\centering
\includegraphics[height=2.2in]{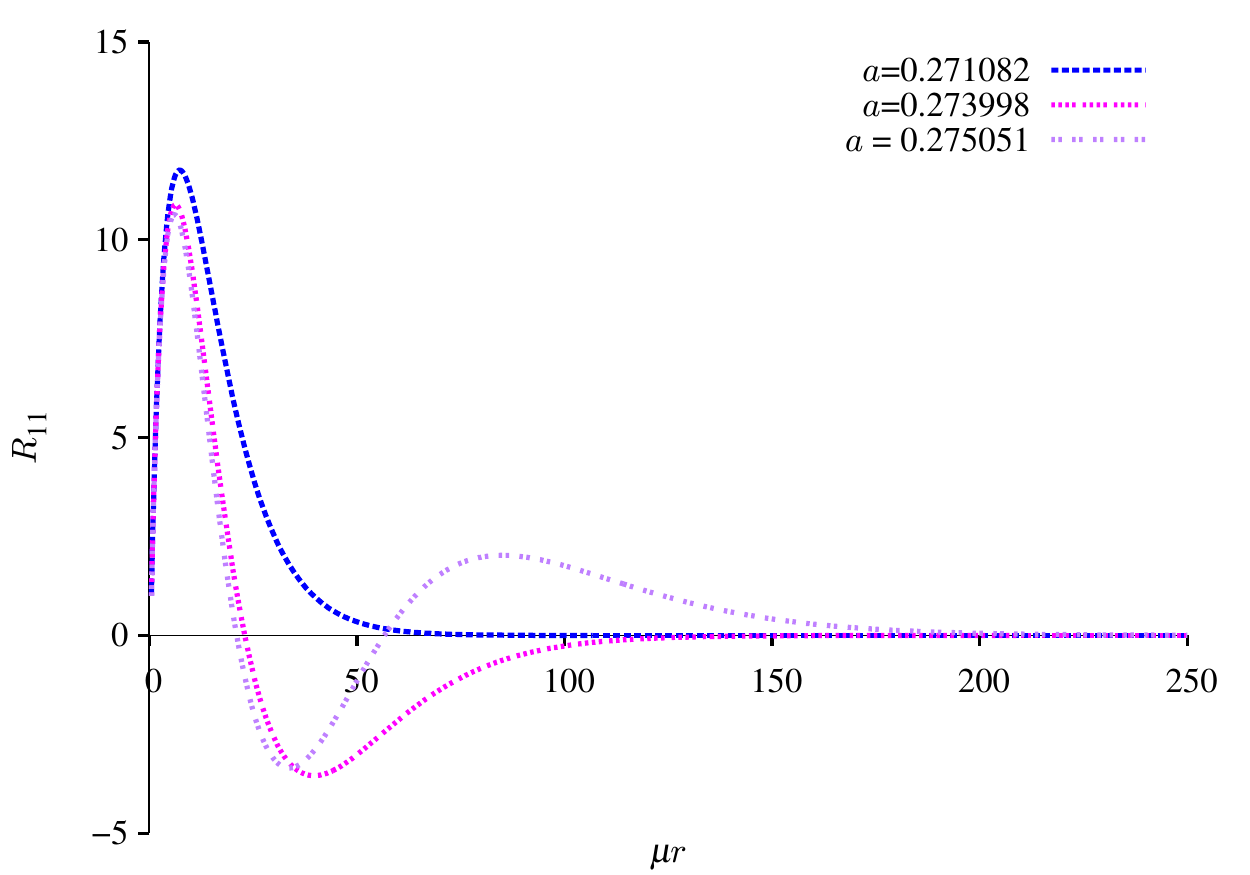} \ \ \ 
\includegraphics[height=2.2in]{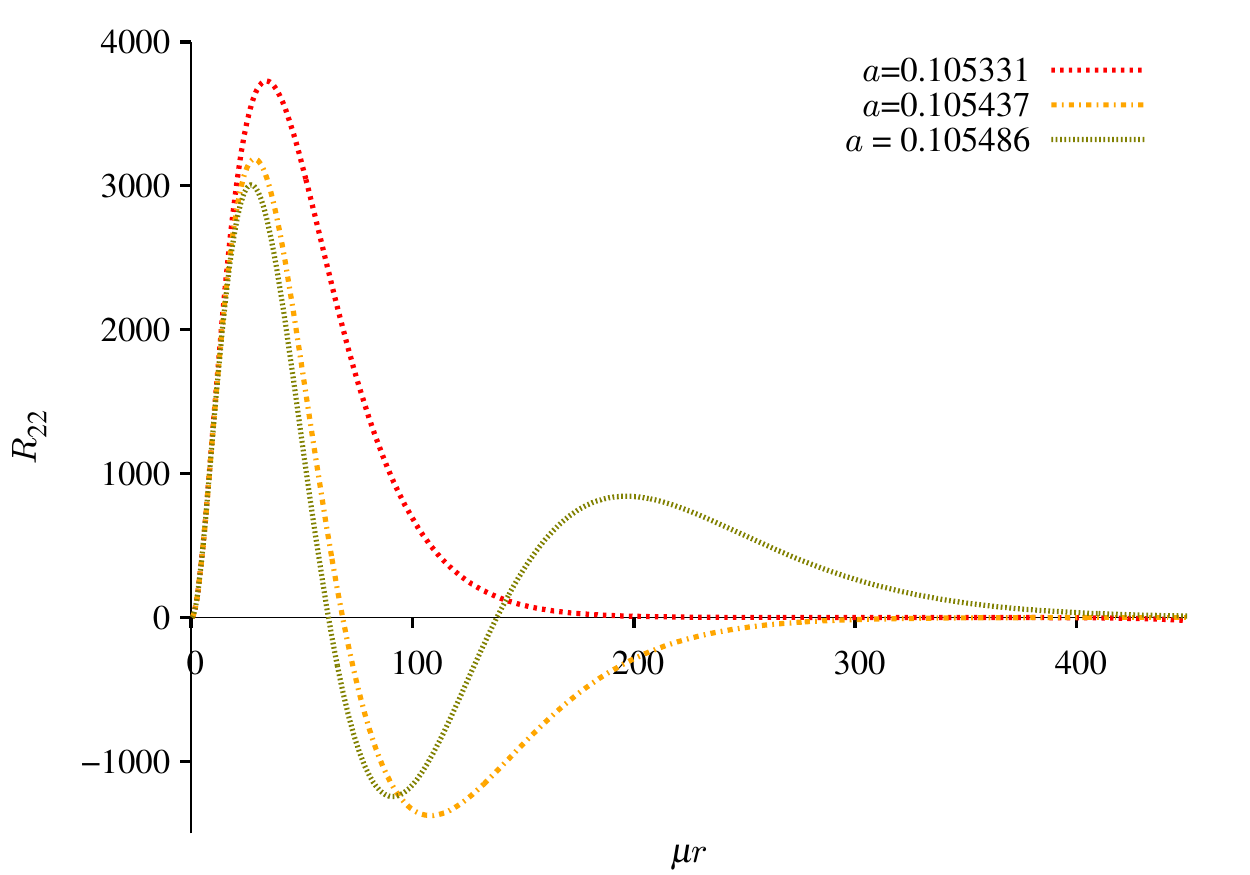}
\caption{Radial solutions $R_{11}$  (left), $R_{22}$  (right) for clouds with $q/\mu=1$ and $n=0,1,2$ in the background of Kerr-Newman with $r_+=0.5$, $Q\mu=0.1$. The corresponding values of $a$ are given in the figure key.} 
\label{rkn}
\end{figure}

\end{widetext}

\begin{figure}[h!]
\centering
\includegraphics[width=\columnwidth]{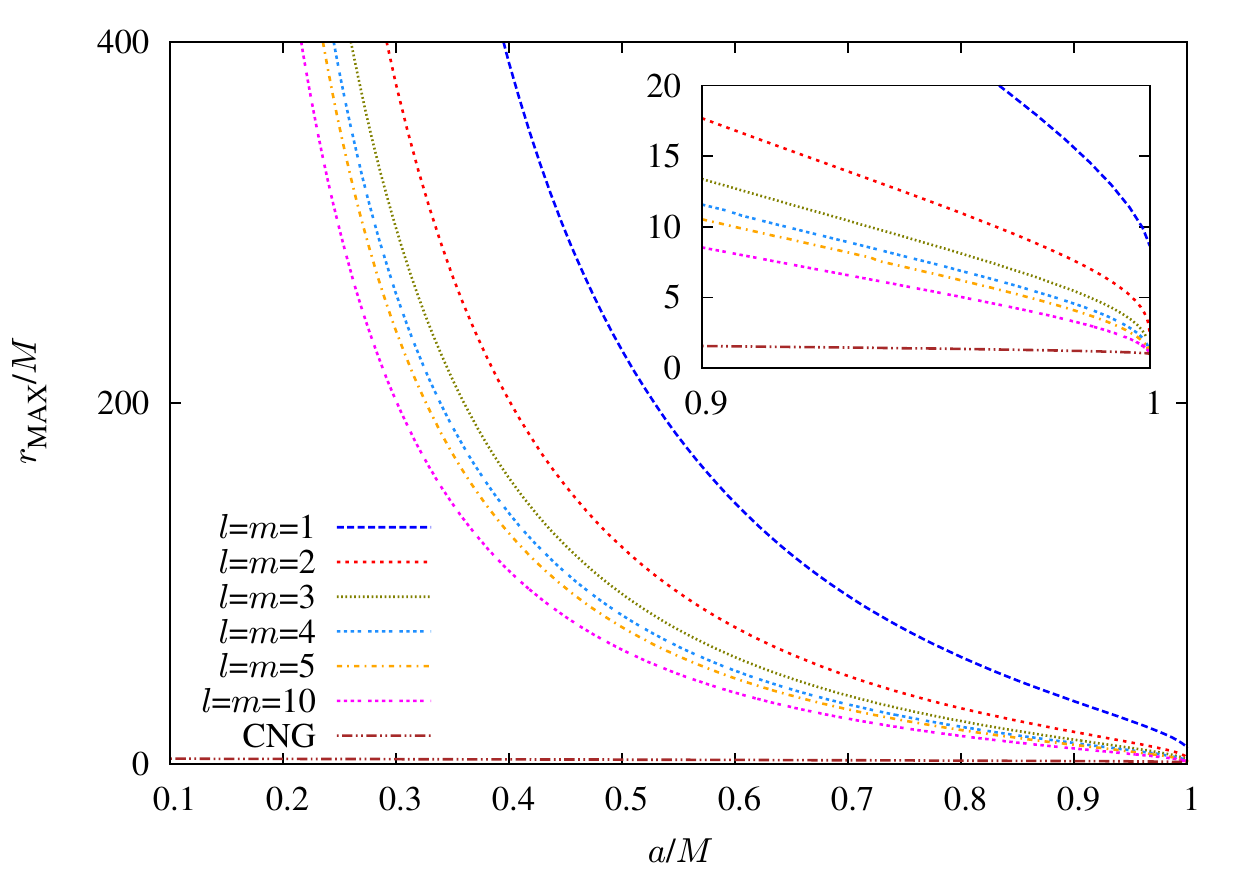}
\caption{``Position" of the clouds, $r_{MAX}/M$ (see definition in the text), and of the co-rotating CNG, as a function of $a/M$ for clouds with $n=0$ and $l=m=1,2,3,4,5,10$ in the Kerr background.}
\label{trm}
\end{figure}

Given the radial profiles, one may ask how close to the horizon the clouds are concentrated. In order to gain some insight into this question, we have plotted in Fig. \ref{trm} the ``position" of the cloud with $l=m=1,2,3,4,5,10$ and $n=0$. By ``position" we mean the value or $r$, denoted $r_{MAX}$, for which the function $4\pi r^2 |R_{lm}|^2$ attains its maximum value, cf. Ref.~\cite{Hod:2012px}. We can see that as $a/M$ decreases, $r_{MAX}/M$ increases, diverging as $a\rightarrow 0$. This behavior is consistent with the fact that Schwarzschild BHs do not support clouds.  As extremality is approached, $a\rightarrow M$, on the other hand, we recover some results by Hod [e.g, for $l=m=1$, $r_{MAX}(a/M=1) = 9.557M=9.557 r_+ $]. It is curious to note that our results for small $l=m$ are in agreement with the `no-short hair' conjecture~\cite{Nunez:1996xv}, which states that, for spherically symmetric BHs, the `hair' should extend beyond  $3 r_+/2$ (which coincides with the position of the circular null geodesic (CNG) for Schwarzschild). But for large $l=m$, the maximum, $r_{MAX}$, approaches the Kerr horizon, as $a\rightarrow M$, in agreement with the behaviour of the co-rotating CNG~\cite{Bardeen:1972fi}, which is also plotted in Fig.~\ref{trm}. The fact that for large $l=m$ the cloud's ``position" can approach arbitrarily close to the horizon was first noted by Hod\footnote{We thank S. Hod for sharing with us this observation and his yet unpublished work~\cite{Hod:2014}.} using the eikonal approximation. These observations support the idea that a more universal measure of the minimal hair extension, valid beyond static BHs, is given by the size of the CNG~\cite{Hod:2011aa}.  

\begin{widetext}

\begin{figure}[h!]
\centering
\includegraphics[height=2.4in]{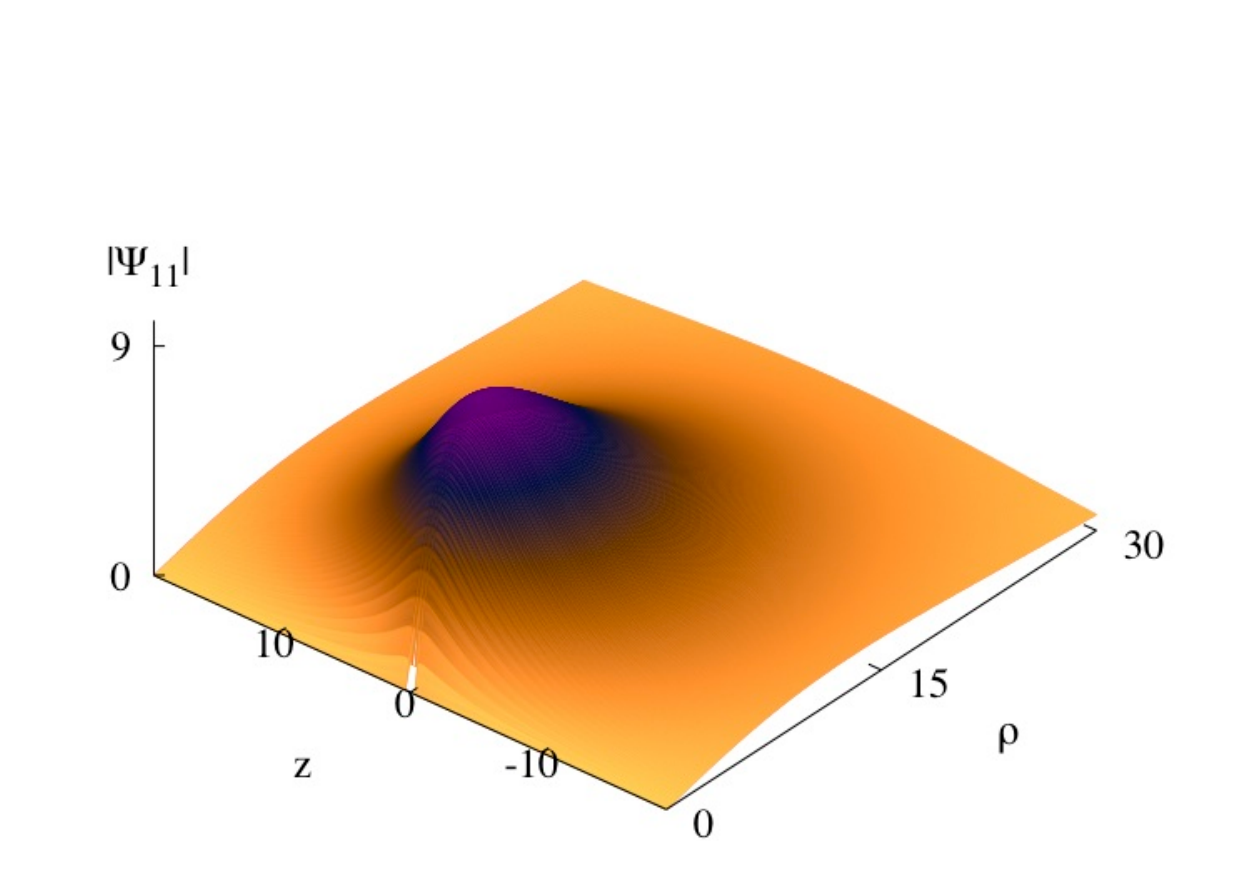} \ \ \ 
\includegraphics[height=2.4in]{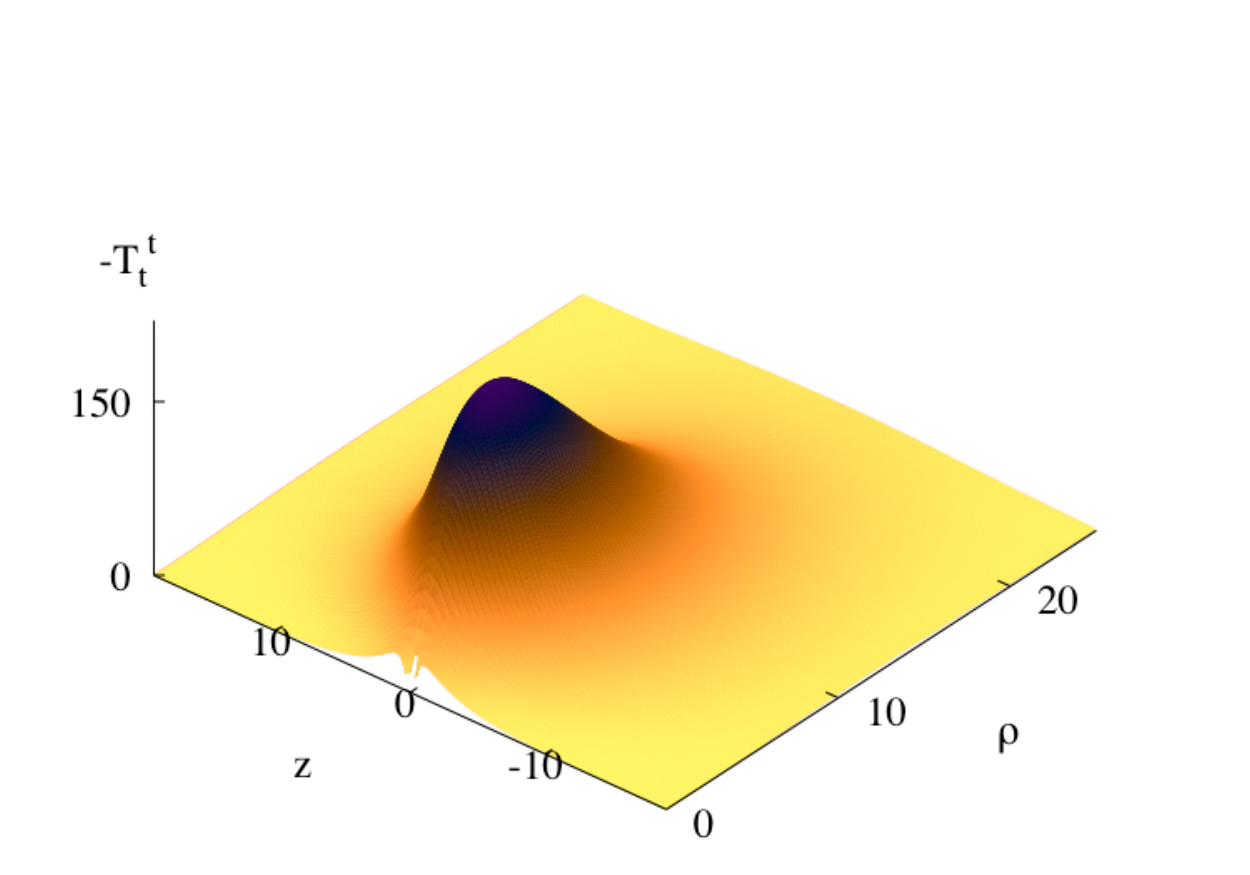}
\caption{Three dimensional spatial distribution of a cloud (left panel) with $n=0$ and $m=l=1$ and its energy density (right panel) in terms of ``polar" coordinates $(\rho,z)$. Both are essentially supported along an equatorial torus, due to the angular distribution of the corresponding spheroidal harmonic.} 
\label{3d}
\end{figure}

\end{widetext}

Finally, in order to have an overview of the full spatial distribution of the clouds and of their energy density, we exhibit in Fig. \ref{3d} a three dimensional plot of both the scalar field distribution (left panel) and the energy density (right panel), for a cloud with $n=0$, $l=m=1$. The particular cloud plotted occurs for background values $r_+=0.43$ and $a=0.26$. For the plot we have normalized the scalar field mode such that $|\Psi_{11}(r=r_H,\theta=\pi/2)|=1$. The plot takes the Boyer-Lindquist coordinates $(r,\theta)$, as standard spherical coordinates and uses the ``polar" coordinates $z=r \cos\theta$ and $\rho=r \sin \theta$. Observe that both the scalar field and the energy density are localized in a toroidal region, well beyond $r_+$. This is expected by virtue of the angular distribution of the corresponding spheroidal harmonic, shown in Fig. \ref{slm3}.  Note that the scalar field vanishes on the $z$-axis. Also, the white space around the origin corresponds to the event horizon (a semi-circle, more clearly seen on the left panel), where the scalar field is non-zero. The energy density plotted is the time-time component of $T^{\ \alpha}_\beta$, where the scalar field energy-momentum tensor is:
\begin{equation}
\label{Tab}
T_{\alpha \beta}= 
2 \Psi_{ , (\alpha}^*\Psi_{,\beta)}-g_{\alpha\beta}  [  \Psi_{,\gamma}^*\Psi^{,\gamma}+\mu^2 \Psi^*\Psi] \ .
\end{equation}

\section{Discussion and Final Remarks}
\label{secfr}
In this paper we have performed a thorough analysis of scalar clouds around Kerr and Kerr-Newman BHs. These configurations are found by solving the massive Klein-Gordon equation, with or without charge, for a test, complex scalar field in the BH background. They are analogous to  atomic orbitals in the sense that there is a scalar field bounded to a central object -- i.e. which decays exponentially asymptotically -- and which is stationary -- i.e. with only a phase-like time dependence. These configurations lie at the threshold of the superradiant instability for a given mode with a set of quantum numbers $(n,l,m)$; fixing them yields a 1-dimensional subset of the 2-dimensional Kerr parameter space or a 2-dimensional subset of the 3-dimensional Kerr-Newman parameter space. To facilitate the analysis of the latter, here we have always examined fixed charge slices of the Kerr-Newman parameter space. Then, for both Kerr and Kerr-Newman cases the clouds are possible along existence lines in the parameter space.

In the Kerr case, the dependence of the existence lines with the quantum numbers $(n,l,m)$ can be summarized as follows. In a BH mass ($M$) vs. BH horizon angular velocity ($\Omega_H$) diagram:
\begin{itemize}
\item[$\bullet$] Nodeless lines $(n=0)$ with $m=l$ are approximately vertical lines and occur for decreasing  values of $\Omega_H$ as the angular quantum numbers $l=m$ increase. Lines with different values of $l=m$ are disconnected. This is in agreement with the naive expectation that the collapse of the cloud is prevented by (stationary) rotation effects and that decreasing the rotation of the BH one must increase the rotation of the cloud and vice-versa. This overall trend had already been observed in \cite{Herdeiro:2014goa}.
\item[$\bullet$] Fixing $m=l$ and increasing the number of nodes $n$, i.e. moving to more excited configurations, the existence line moves to slightly higher values of $\Omega_H$. All these lines are connected: they converge when the BH mass tends to zero. Again, a naive interpretation is that clouds with nodes are excited states, hence more energetic and thus require a larger background rotation for equilibrium. Their ``weight" however, becomes irrelevant as the background mass vanishes which agrees with the convergence of these existence lines in the $M\rightarrow 0$ limit.
\item[$\bullet$] Fixing $m$ and $n$ and increasing $l$, again the existence line moves to slightly higher values of $\Omega_H$, and again all these lines are connected as the BH mass tends to zero.  This overall trend was also briefly mentioned in \cite{Herdeiro:2014goa}.
\end{itemize}

Adding charge to both the background and the field introduces two qualitatively new effects in the same type of diagram as before:
\begin{itemize}
\item[$\bullet$]  When the background charge and the field charge have the same (opposite) sign, the existence line for a given set of quantum numbers moves to lower (higher) values of $\Omega_H$, for fixed $M$. Again, this is in agreement with the naive expectation that there is now Coulomb repulsion (attraction) between the background BH and the cloud that needs to be  balanced by smaller (higher) background rotation. 
\item[$\bullet$] The existence lines stop being essentially vertical. The reason is that they cannot approach the $M\rightarrow 0$ limit, since there is a minimal BH mass for a given BH charge $Q$ which still allows the existence of an event horizon. In the $Qq>0$ case, fixing the field charge equal to the field mass, one observes that as the minimal mass is approached, then $\Omega_H\rightarrow 0$ and $M\rightarrow Q$. The configuration approached is a marginal charged cloud, in the nomenclature of Ref.~\cite{Sampaio:2014swa}.
\end{itemize}

Our numerical results for the existence lines were also compared with  some analytic approximations found in the literature, which in principle are valid for either small rotation or high rotation. Somewhat surprisingly, in the cases shown here, these approximations yield a fairly accurate estimate even far away from their \textit{a priori} validity region. 

We have also described the spatial distribution of the clouds. A full picture is obtained by describing the angular and radial dependence separately. 
We have given examples of both spheroidal harmonics 
and of radial functions. An interesting property of the latter is that an appropriately defined radial ``position" for the clouds increases as the rotation of the background decreases. Thus the smaller radial position is obtained for extremal BHs and it is in agreement with a generalization of the `no-short hair' conjecture suggested by Hod~\cite{Hod:2011aa}.

 As already  observed, cloud solutions  can be taken 
 as a smoking gun for the existence of Kerr-(Newman) BHs with scalar hair, as fully non-linear solutions of the Einstein-Klein-Gordon(-Maxwell) system
 \cite{Herdeiro:2014goa}. But that does not imply that all hairy BHs are revealed by such a test field analysis. A remarkable example concerns Myers-Perry BHs, which can support a scalar hair which relies on non-linear effects~\cite{Brihaye:2014nba}.

The analysis in this work was restricted
to the simple case of scalar fields on Kerr(-Newman) BHs,  for which the wave equation separates.
Similar results are expected to hold as well for other 
rotating BH 
backgrounds afflicted by superradiant instabilities, as well as other fields that may trigger such instabilities. In all such cases scalar (or other spin fields) clouds should occur at the threshold of the
superradiant instability. One particularly interesting case that we expect is the existence of Proca clouds around Kerr BHs (see \cite{Sampaio:2014swa} for the study of Proca quasi-bound states around Schwarzschild BHs).

We would also like to comment on the stability of the clouds discussed in this work. Since they only exist along lines in the Kerr 2-dimensional parameter space, one may expect that the clouds are unstable solutions. An argument that they are actually dynamical attractors is the following. First observe that  $\omega/m$ defines an angular velocity for the cloud. The bound state condition   $\omega=\omega_c=m\Omega_H$, can then be interpreted as the angular velocity of the cloud being synchronous with that of the BH horizon. Now let us fix a Kerr background with a given horizon angular velocity $\Omega_H$ and consider a quasi-bound state with a frequency $\mathcal{R}(\omega)$ slightly smaller (larger) than $\omega_c$. The quasi-bound state will be in the superradiant (decaying) regime. Thus it will be amplified (absorbed) by the BH. Following the evolution of the background BH in a quasi-static approximation, the BH loses (gains) mass and angular momentum and its angular velocity decreases (increases). The process only stops when the horizon angular velocity of the evolving BH approaches $\Omega_H\rightarrow \mathcal{R}(\omega)/m$, at which point the imaginary part of the quasi-bound state frequency approaches zero. Thus it seems plausible that clouds are dynamical equilibrium configurations; in other words, that dynamics wants to lock quasi-bound states in synchronous rotation with the BH. This argument is reminiscent of the synchronization of orbital and rotation periods of astronomical bodies (like the Moon-Earth system) due to tidal effects and friction. This analogy supports the idea of a connection between tidal acceleration and superradiant scattering around spinning BHs~\cite{Cardoso:2012zn}.

  Finally,  a more involved picture is found when turning on a suitable
self-interaction potential of the scalar field.
Then the (non-linear) Klein-Gordon equation
possesses bound state solutions already in a flat spacetime background -- {\it the Q-balls}~
\cite{Coleman:1985ki,Radu:2008pp}.
Remarkably,  spinning  Q-balls
survive when replacing the Minkowski background with a Kerr metric, provided
the relation (\ref{cond})
connecting the scalar field frequency and the BH
event horizon velocity
is satisfied
(note that the variables do not separate in this case).
The resulting solutions exhibit a more complicated pattern than the clouds presented herein, 
covering
a compact region of the 2-dimensional parameter space of the Kerr BHs, rather than existence lines.
These solutions are reported elsewhere~\cite{Herdeiro:2014pka}. 

\section*{Acknowledgements}
We thank Helgi R\'unarsson and Juan Carlos Degollado for discussions on this subject. The authors are partially supported by the FCT IF program. The work in this paper is also supported by the grants PTDC/FIS/116625/2010 and  NRHEP--295189-FP7-PEOPLE-2011-IRSES. The authors would like also to thank Conselho Nacional de Desenvolvimento Cient\'ifico e Tecnol\'ogico (CNPq), Coordena\c{c}\~ao de Aperfei\c{c}oamento de Pessoal de N\'ivel Superior (CAPES), and Funda\c{c}\~ao Amaz\^onia Paraense de Amparo \`a Pesquisa (FAPESPA), from Brazil, for partial financial support.

\newpage

\bibliographystyle{h-physrev4}
\bibliography{clouds}

\end{document}